\documentclass[twocolumn,prl,amsmath,amssymb]{revtex4}
\usepackage{graphicx}% Include figure files
\usepackage{dcolumn}% Align table columns on decimal point
\usepackage{bm}% bold math

\begin{document}

\preprint{APS/123-QED}

\title{Vortex solitons in photonic crystal fibers}% Force line breaks with \\

\author{Albert Ferrando and Mario Zacar\'es}
 \affiliation{Departament d'\`Optica, Universitat de Val\`encia. Dr. Moliner, 50. E-46100 Burjassot (Val\`encia), Spain}%Lines break automatically or can be forced with \\
\author{Pedro Fern\'andez de C\'ordoba}%
\affiliation{
Departamento de Matem\'atica Aplicada, Universidad Polit\'ecnica de Valencia. Camino de Vera, s/n. E-46022 Valencia, Spain
}%

\author{Daniele Binosi}
\affiliation{
Departament de F\'{\i}sica Te\`orica, Universitat de Val\`encia. Dr. Moliner, 50. E-46100 Burjassot (Val\`encia), Spain
}%

\author{Juan A. Monsoriu}
\affiliation{
Departamento de F\'{\i}sica  Aplicada, Universidad Polit\'ecnica de Valencia. Camino de Vera, s/n. E-46022 Valencia, Spain
}
\date{\today}% It is always \today, today,

\begin{abstract}
We demonstrate the existence of vortex soliton solutions
in photonic crystal fibers. We analyze the role played by the photonic
crystal fiber defect in the generation of optical vortices. An analytical
prediction for the angular dependence of the amplitude and phase of
the vortex solution based on group theory is also provided. Furthermore,
all the analysis is performed in the non-paraxial regime. 
\end{abstract}
%\ocis{060.4370, 190.4379, 190.0190}
% REPLACE WITH CORRECT OCIS CODES FOR YOUR ARTICLE
% NOTE: \ocis{} IS ALIASED TO \pacs{} BUT MUST
% FORMAT THE TERMS CORRECTLY FOR EACH JOURNAL
%\ocis{(060.4370) Nonlinear optics, fibers; (190.4370) Nonlinear optics, fibers.}
\maketitle %% NULL FUNCTION WITH LATEX 2e

Spatial soliton solutions in 1D periodic structures have been theoretically
predicted \cite{christodoulides-ol13_794} and experimentally demonstrated
in nonlinear waveguide arrays \cite{eisenberg-prl81_3383}. More recently,
similar structures have been observed in optically induced gratings
\cite{fleischer-prl90_023902,neshev-ol28__710} and, for the first
time, in 2D photonic lattices \cite{fleischer-nature422_147}. Modeling
has been traditionally performed using the discrete nonlinear Schr\"{o}dinger
equation (NLSE)\cite{christodoulides-ol13_794,kivshar-ol18_1147},
valid only in the so-called tight-binding approximation. However,
the accurate study of nonlinear solutions in optically-induced lattices
requires the resolution of the NLSE with a continuous periodic potential
model \cite{fleischer-prl90_023902,fleischer-nature422_147,neshev-ol28__710}.
Discrete optical vortices have been also predicted in the discrete
NLSE \cite{malomed-pre64_026601} and in continuous models \cite{yang-arXiv:physics/0304047}
and, recently, experimentally observed in optically-induced 2D lattices
\cite{neshev-arXiv:nlin/0309018}.

\begin{figure}[h]\centerline{\scalebox{1}{\includegraphics{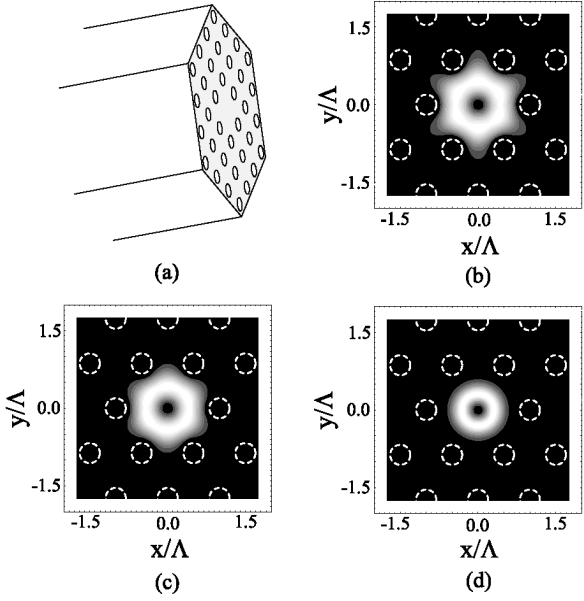}}}
 \caption{(a) Schematical representation of a PCF. 
(b)-(d) Amplitudes of several
   vortices for increasing values of $\gamma$ in a PCF with  $\Lambda=31\,\mu\mathrm{m}$ and  $a=6\,\mu\mathrm{m}$ ($\lambda=1064\,\mathrm{nm}$): (b) $\gamma=0,45\times 10^{-3}$;
(c) $\gamma=0,95\times 10^{-3}$; (d) $\gamma=1,75\times 10^{-3}$.
 \label{fig:(a)-PCF scheme.(b) Optical vortices}}
 \end{figure}

A different scenario appears in a 2D photonic crystal in the presence
of a defect. An example of such a system is a photonic crystal fiber
(PCF). A PCF is a silica fiber that possesses a regular array of air-holes
extending along the entire fiber length (see Fig.\ref{fig:(a)-PCF scheme.(b) Optical vortices}(a)).
It is an axially invariant 2D photonic crystal with a central defect
(the PCF core, where guidance occurs). It has been recently proven
that a PCF acting as a 2D nonlinear photonic crystal can support and
stabilize fundamental solitons \cite{ferrando-oe11_452}.
Unlike fundamental solitons embedded in perfectly periodic potentials
\cite{fleischer-nature422_147,yang-arXiv:physics/0304047}, a distinguishing
feature of these systems is that no power gap is needed to generate
the nonlinear localized solution. In this letter we go a step further
and give numerical evidence of the existence of vortex solitons
in PCF's.

When one considers that silica can have a nonlinear response, electromagnetic
propagation in a PCF is given by the following nonparaxial equation
($\nabla\cdot\mathbf{E}\approx0$):

\begin{equation}
[\nabla_{t}^{2}+k_{0}^{2}(n_{0}^{2}(\vec{x})+n_{2}^{2}(\vec{x})|\mathbf{E}|^{2})]\mathbf{E}=-\partial^{2}\mathbf{E}/\partial z^{2}, \label{eq_evolucio}
\end{equation}

$\nabla_{t}^{2}$ being the 2D-transverse Laplacian operator and $k_{0}=\omega_{0}/c$
the vacuum wavenumber. The linear refractive index profile function
$n_{0}(x,y)$ is 1 in the air-holes and equals $n_{\mathrm{silica}}$
in silica, whereas the nonlinear index profile function $n_{2}(x,y)$
is different from zero only in silica ($n_{2(\mathrm{silica})}^{2}\equiv3\chi_{(\mathrm{silica})}^{(3)}/(2\varepsilon_{0}cn_{0(\mathrm{silica})})$).
We search for monochromatic (or quasi-monochromatic) stationary electric
field solutions with well-defined constant polarization: $\mathbf{E}(x,y,z,t)\approx\mathbf{u}\phi(x,y)e^{i(\beta z-\omega_{0}t)}$.
That is, we analyze the following equation:

\begin{equation}
L(|\phi|)\phi=\beta^{2}\phi\,\,\,\,\,\,\,\,\,\,\, L(|\phi|)=L_{0}+L_{NL}(|\phi|),\label{eq:Lbeta}
\end{equation}

 where $L_{0}=\nabla_{t}^{2}+k_{0}^{2}n_{0}^{2}(x,y)$ and $L_{NL}(|\phi|)=k_{0}^{2}n_{2}^{2}(x,y)|\phi|^{2}$
stand for the linear and nonlinear parts of the differential operator
$L(|\phi|)$, respectively. Due to the geometry of the air-holes distribution,
both $n_{0}(x,y)$ and $n_{2}(x,y)$ are invariant under the action
of the $\mathcal{C}_{6v}$ point-symmetry group. This group is constituted
by discrete $\pi/3$-rotations ($R_{\pi/3}$) plus specular reflections
with respect to the $x$ and $y$ axes: $y\stackrel{R_{x}}{\rightarrow}-y$
and $x\stackrel{R_{y}}{\rightarrow}-x$, or, equivalently, $\theta\stackrel{R_{x}}{\rightarrow}-\theta$
and $\theta\stackrel{R_{y}}{\rightarrow}\pi-\theta$, in polar coordinates.

Our aim is to look for nonlinear solutions of Eq.(\ref{eq:Lbeta})
beyond the fundamental solitons reported in Ref.\cite{ferrando-oe11_452}.
Our search strategy will be based on group theory arguments, which
will be enlightening in the classification and understanding of new
solutions according to the sixth-fold symmetry of the system. In the
linear case, this approach has been adopted to study the degeneracy
of solutions in a PCF \cite{steel-ol26_488}. Here, we will use it
to obtain analytical expressions for the angular dependency of solutions. 
According to group
theory, since $L_{0}$ is invariant under the $\mathcal{C}_{6v}$
group, all its eigenfunctions have to lie on discrete
representations of this group \cite{hamermesh64}. Solutions of the
vortex type belong to one of the two two-dimensional representation
of $\mathcal{C}_{6v}$. A pair of functions $(\phi_{l},\phi_{l}^{*}$)
($l=1,2$) belonging to these representations have the following transformation
properties: $(\phi_{l},\phi_{l}^{*})\stackrel{R_{\pi/3}}{\rightarrow}(\epsilon^{l}\phi_{l},\epsilon^{*l}\phi_{l}^{*})$
$\epsilon=e^{i\pi/3}$, $(\phi_{l},\phi_{l}^{*})\stackrel{R_{x}}{\rightarrow}(\phi_{l}^{*},\phi_{l})$
and $(\phi_{l},\phi_{l}^{*})\stackrel{R_{y}}{\rightarrow}(-1)^{l}(\phi_{l}^{*},\phi_{l})$.
These symmetry constraints impose the form of these functions:

\begin{equation}
\phi_{l}=r^{l}e^{il\theta}\phi_{l}^{s}(r,\theta)\exp[i\phi_{l}^{p}(r,\theta)]\,\,\, l=1,2,\label{eq:ansatz}\end{equation}

 where $\phi^{s}(r,\theta)$ is a scalar function, characterized by
$\phi^{s}(r,\theta+\pi/3)=\phi^{s}(r,\theta)$ and $\phi^{s}(r,-\theta)=\phi^{s}(r,\pi-\theta)=\phi^{s}(r,\theta)$,
and $\phi^{p}(r,\theta)$ is a pseudoescalar function characterized
by $\phi^{p}(r,\theta+\pi/3)=\phi^{p}(r,\theta)$ and $\phi^{p}(r,-\theta)=\phi^{p}(r,\pi-\theta)=-\phi^{p}(r,\theta)$.
Both $\phi^{s}$ and $\phi^{p}$ are periodic functions of $\theta$,
so they can be expanded in Fourier series in $\cos(6n\theta)$ and
$\sin(6n\theta)$ ($n\in N$). Reflection symmetry forces
$\phi^{s}$($\phi^{p}$) to depend on the cosine (sine) terms only:
$\phi_{l}^{s}(r,\theta)=\sum_{n}a_{ln}(r)\cos(6n\theta)$ and $\phi_{l}^{p}(r,\theta)=\sum_{n}b_{ln}(r)\sin(6n\theta)$. 

In the nonlinear case, functions $\phi_{l}$ of the form given by
Eq.(\ref{eq:ansatz}) can satisfy self-consistency, according to group
theory. The full operator $L(|\phi_{l}|)=L_{0}(n_{0})+L_{NL}(r^{l}\phi_{l}^{s};n_{2})$
is invariant under the $\mathcal{C}_{6v}$ group because $L_{0}$,
$r^{l}$ , $\phi_{l}^{s}$ and $n_{2}$ are all $\mathcal{C}_{6v}$
invariants. Thus $L(|\phi_{l}|)$, like in the linear case, provides
the representation where $\phi_{l}$ lies on. However, group self-consistency
is not sufficient to ensure that the resolution of the nonlinear equation
(\ref{eq:Lbeta}) with the ansatz (\ref{eq:ansatz}) provides a nontrivial
solution in all cases, since Eq.(\ref{eq:Lbeta}) always admits the
$\phi=0$ solution. We solve Eq.(\ref{eq:Lbeta}) by means of the Fourier
iterative method previously used in Ref.\cite{ferrando-oe11_452}
to find soliton solutions in PCF's. These solutions also fulfill
the group self-consistency condition and they belong to the fundamental
representation of \emph{$\mathcal{C}_{6v}$}. The important
difference now is that we restrict ourselves to a different representation
space; we search for nonlinear solutions in the $l=1,2$ representations
of $\mathcal{C}_{6v}$. The $\phi_{l}$ solution and its conjugate
$\phi_{l}^{*}$ represent a vortex and an anti-vortex soliton of order
$l$, respectively.

\begin{figure}[h]\centerline{\scalebox{1}{\includegraphics{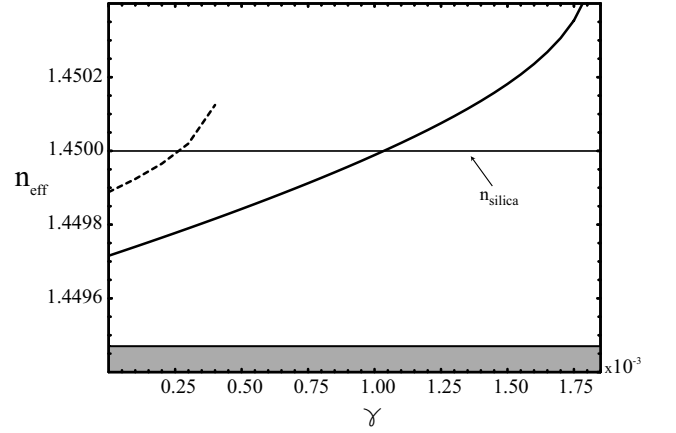}}}
\caption{Effective index of a family of vortex solutions as a function
of $\gamma$ (solid line). Same for a family of fundamental solitons (dashed
line), as in Ref.\cite{ferrando-oe11_452}. The shadow region corresponds
to the conduction band, constituted by Bloch modes, of
the 2D photonic cladding with $\Lambda=31\,\mu\mathrm{m}$ 
and $a=10\,\mu\mathrm{m}$
($\lambda=1064\,\mathrm{nm}$).
\label{fig:effective-index}}
\end{figure}

We have simulated large-scale PCF's (in order to prevent silica breakdown
\cite{ferrando-oe11_452}) with a lattice period, or pitch,
$\Lambda=31\,\mu\mathrm{m}$, and several air-hole radius at a fixed wavelength $\lambda=1064\,\mathrm{nm}$ (a convenient
quasi-continuum source). We define a dimensionless nonlinear parameter
as $\gamma=n_{2(\mathrm{silica})}^{2}P/A_{0}$ ($A_{0}$ is an area
parameter: $A_{0}=\pi(\Lambda/2)^{2}$; $P$ is the total power carried
by the solution) and perform calculations for increasing values of
$\gamma$. We discover that a family of optical vortex solutions of
the type given by the ansatz (\ref{eq:ansatz}) is found. The amplitudes
of several solutions (with $l=1)$ for increasing values of $\gamma$
are represented in Fig.\ref{fig:(a)-PCF scheme.(b) Optical vortices}(b)-(d).
According to group theory (Eq.(\ref{eq:ansatz})), these amplitudes
have to be scalar functions, thus showing full invariance under \emph{$\mathcal{C}_{6v}$}, and they have to vanish at
the origin, as seen in Fig.\ref{fig:(a)-PCF scheme.(b) Optical vortices}(b)-(d).
Besides, they become gradually narrower as the nonlinearity increases.
It is interesting to plot the effective index of these solutions ($n_{\mathrm{eff}}=\beta/k_{0}$)
versus $\gamma$, as shown in Fig.\ref{fig:effective-index}. The
value of $n_{\mathrm{eff}}$ increases as $\gamma$ increases, accordingly
to the narrowing of solutions depicted in Fig.\ref{fig:(a)-PCF scheme.(b) Optical vortices}(b)-(d).
For comparison, we have also included the $n_{\mathrm{eff}}(\gamma)$
curve of the fundamental soliton family for the same PCF structure, as
in Ref.\cite{ferrando-oe11_452}. Both lie on the upper forbidden
band of the perfectly periodic cladding structure. Here it can be
clearly envisaged the role of the central defect. Notice that there
is no threshold $\gamma$ (power) to generate a nonlinear vortex soliton,
i.e, there is a continuum of solutions in $\gamma$ starting from
the linear (TE or TM) mode. Unlike in perfectly periodic structures,
the defect can eliminate the presence of a threshold power \cite{yang-arXiv:physics/0304047}.
If we consider structures where the linear mode is not present \cite{ferrando-josaa17_1333},
a threshold $\gamma$ is necessary to generate the nonlinear solution
in the forbidden band. By playing with the geometric parameters
of the PCF, it is possible to tune this threshold value. In all cases,
this value is much lesser than in perfectly periodic structures, which
can be important for experimental purposes.

\begin{figure}[h]\centerline{\scalebox{1}{\includegraphics{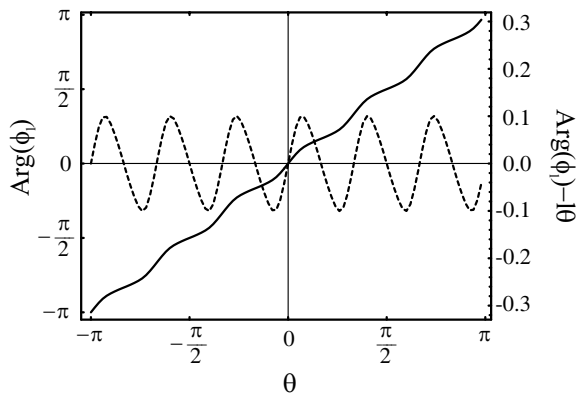}}}
\caption{Phase of a vortex with $l=1$ 
at $r=21\,\mu\mathrm{m}$.
We represent both the total phase $\mathrm{arg}(\phi_{l})$ (solid
line) and the group phase $\mathrm{arg}(\phi_{l})-l\theta$ (dashed
line).
\label{fig:vortex-phase}}
\end{figure}

Even more interesting is the phase behavior of these solutions. We
show in Fig.\ref{fig:vortex-phase} a typical phase profile of a PCF
vortex calculated at a fixed radius. It shows strict accordance
with the group theory prediction for the phase: $\mathrm{arg}(\phi_{l})=l\theta+\phi_{l}^{p}(r,\theta)=l\theta+\sum_{n=1}^{\infty}b_{ln}(r)\sin(6n\theta)$.
The phase of these vortex solutions differs from that of an ordinary
vortex by the presence of the pseudo-scalar function extra term: the
group phase. Besides the existence of the standard linear behavior
in $\theta$, the group phase provides an additional sinusoidal dependence
on the angle (with period determined by the group order: $\pi/3$
for order 6) not present in ordinary vortices. 

\begin{figure}[h]\centerline{\scalebox{1}{\includegraphics{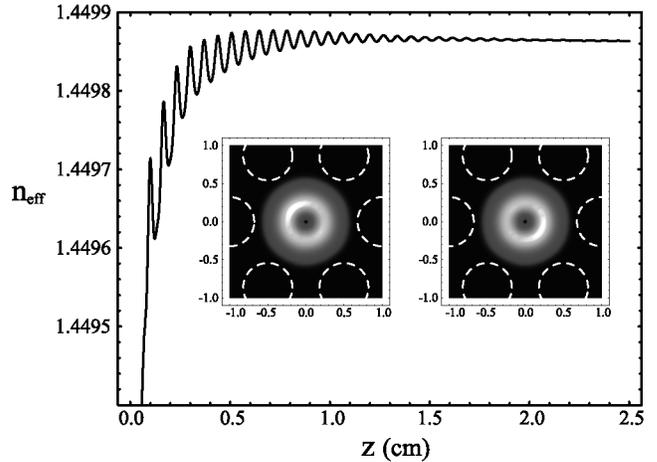}}}
\caption{Axial evolution of 
$\left\langle n_{\mathrm{eff}}(z)\right\rangle $
for a diagonal perturbation. Inset: two snapshots of the asymptotic behavior of a non-diagonal oscillation instability (notice the different positions of the peak intensity).
\label{fig:axial-evolution}}
\end{figure}

In order to check the stability of a vortex soliton solution $\phi_{l}$,
we have to consider $z$-dependent perturbations. This implies solving
the non-paraxial equation (\ref{eq_evolucio}) for the perturbed field
$\phi'=\phi_{l}+\delta\phi$. In terms of group theory, ($\phi_{l},\phi_{l}^{*}$)
constitute a basis of the $l$ representation of $\mathcal{C}_{6v}$,
so we can consider two types of perturbations around a vortex solution
$\phi_{l}$: diagonal perturbations, in which $\phi'$ remains in
the one-dimensional subspace spawned by $\phi_{l}$, and non-diagonal,
where the perturbation takes $\phi'$ out of this subspace. An initial
condition of the form $\phi'=re^{i\theta}\exp[-(r/r_{0})^2]$
is an example of diagonal perturbation for a vortex ($l=1$). Non-paraxial
evolution shows that a vortex is stable under diagonal perturbations.
In Fig.\ref{fig:axial-evolution}, we show the stabilization of $\left\langle n_{\mathrm{eff}}(z)\right\rangle \equiv\left(\int\phi^{*}L(\phi)\phi\right)^{1/2}/k_{0}$
to the $n_{\mathrm{eff}}$ of an asymptotic vortex state. Diagonal
perturbations generalize the concept of radial perturbations to the
discrete symmetry case. We have also simulated non-diagonal perturbations.
In this case, an oscillatory instability occurs, as seen in Fig.\ref{fig:axial-evolution}(inset).
Non-diagonal perturbations correspond to azimuthal ones in
the radially symmetric case. Despite the presence of azimuthal instabilities,
no collapse is detected due to the non-paraxial nature of evolution
\cite{akhmediev-ol18_411}.

Summarizing, we propose a novel way of generating vortex solitons,
based on PCF's, in which the presence of the defect plays a crucial
role. Our analytical approach, based on group theory,
is completely general and applies to any system owning a discrete
invariance, no matter is periodic or non-periodic, linear or nonlinear.
Finally, the use of a non-paraxial equation permits to analyze new
phenomena beyond the paraxial approximation. 

We thank H. Michinel for useful discussions. This
work was financially supported by the Plan Nacional I+D+I (grant TIC2002-04527-C02-02),
Ministerio de Ciencia y Tecnolog\'{\i}a (Spain) and FEDER funds.
M. Zacar\'{e}s acknowledges Fundaci\'{o}n Ram\'{o}n
Areces grant.

\end{document}